\documentclass{dhbenelux}
\usepackage[doublespacing]{setspace}
\usepackage{booktabs} 
\usepackage{authblk} 
\usepackage{graphicx} 
\setcitestyle{apa}

\DeclareOldFontCommand{\bf}{\normalfont\bfseries}{\mathbf}
\author[1]{Harlan Campbell\footnote{Please contact: harlan.campbell@stat.ubc.ca}}
\author[1]{Paul Gustafson }

\affil[1]{University of British Columbia}

\title{re:Linde et al. (2021): The Bayes factor, HDI-ROPE and frequentist equivalence tests can all be reverse engineered -almost exactly- from one another}

\begin{document}

\maketitle

\vspace{-1.85cm}

\abstract{{\footnotesize{ABSTRACT -  Following an extensive simulation study comparing the operating characteristics of three different procedures used for establishing equivalence (the frequentist `TOST'', the Bayesian ``HDI-ROPE'', and the Bayes factor interval null procedure), Linde et al. (2021) conclude with the recommendation that ``researchers rely more on the Bayes factor interval null approach for quantifying evidence for equivalence.''    We redo the simulation study of Linde et al. (2021) in its entirety but with the different procedures calibrated to have the same predetermined maximum type 1 error rate.  Our results suggest that, when calibrated in this way,  the Bayes Factor, HDI-ROPE, and frequentist equivalence tests all have similar -almost exactly- type 2 error rates.  In general any advocating for frequentist testing as better or worse than Bayesian testing in terms of empirical findings seems dubious at best. If one decides on which underlying principle to subscribe to in tackling a given problem, then the method follows naturally.  Bearing in mind that each procedure can be reverse-engineered from the others (at least approximately), trying to use empirical performance to argue for one approach over another seems like tilting at windmills.}}
}

\vspace{0.1cm}

{\footnotesize{\paragraph{Acknowledgments - }  Many thanks the authors of \citet{linde2020decisions} who provided open access to their well documented code without which our work would not have been possible.}}

{\footnotesize{\paragraph{Code - }  Code to replicate our simulation study and all of the Figures is available at \url{https://github.com/harlanhappydog/reLinde}.}}



\pagebreak

\section{Introduction}

\citet{linde2020decisions} describe and compare three different approaches for finding evidence of equivalence between two groups:

\begin{enumerate}
    \item  ``TOST'': the frequentist two one-sided t-tests procedure with $\alpha=0.05$ \citep{hodges1954testing, schuirmann1987comparison, westlake1976symmetrical};
    
    \item ``HDI-ROPE'': the Bayesian highest density interval (HDI) region of practical equivalence procedure with a 95\% HDI \citep{kruschke2011bayesian, kruschke2013bayesian}; and
    
    \item ``BF'': the Bayes factor interval null procedure with one of  two different BF decision thresholds, either $BF_{thr} = 3$ or $BF_{thr} = 10$; see \citet{morey2011bayes}.
\end{enumerate}

Following an extensive simulation study, \citet{linde2020decisions} conclude with the recommendation that ``researchers rely more on the Bayes factor interval null approach for quantifying evidence for equivalence.''  This recommendation is based on the finding that the TOST and HDI-ROPE have ``limited discrimination capabilities when the sample size is relatively small.''  However, we suspect that the same remark could be made about the BF procedure if it were calibrated so as to maintain a predetermined maximum type 1 error rate.

Motivated by this suspicion, we repeat the simulation study of \citet{linde2020decisions} in its entirety to determine how the different methods compare when they are calibrated to all have the same maximum type 1 error rate. \citet{linde2020decisions} write: ``In general, it is important to evaluate statistical testing approaches based on both types of errors,'' i.e., both the type 1 error and the type 2 error.  However, the degree of type 1 error is dependent on the degree of type 2 error and vice-versa.  Therefore, in order to evaluate and compare the statistical power of different tests on a level playing field, one must proceed by first calibrating each test to have the same predetermined maximum type 1 error rate.

\section{Methods}
Our simulation study is identical to the one conducted by \citet{linde2020decisions} with a few notable exceptions.

First, for frequentist equivalence testing we  consider the so-called ``optimal test'' based on the folded-Normal distribution \citep{romano2005optimal} in addition to the TOST.   It is ``well known in the mathematical statistics literature'' \citep{mollenhoff2019efficient} that the optimal test is more powerful than the TOST, particularly for small sample sizes. While the TOST procedure for frequentist equivalence testing works well for moderate and large sample sizes, it is sub-optimal for small sample sizes; see \cite{lehmann2006testing} and \cite{wellek2010testing}.  Note that both tests are asymptotically equivalent (i.e., essentially identical for sufficiently large sample sizes).

The optimal test can be summarized as follows.  Reject the null hypothesis ($H_{0}: |\delta| \ge m$), whenever:
\begin{equation}
    |\bar{X}_{1} - \bar{X}_{2}| < u_{\alpha}
\end{equation}
\noindent where $\delta$ is the true difference in group means, $m$ is the equivalence margin, $\bar{X}_{1}$ is the observed sample mean for the first group, $\bar{X}_{2}$ is the observed sample mean for the second group, and $u_{\alpha}$ is the $\alpha$-quantile of the folded Normal distribution, $N_{F}(m, \hat{\sigma}_{P}^{2})$, with location parameter equal to the margin, $m$, and scale parameter equal to estimated pooled variance of the data, $\hat{\sigma}_{P}^{2}$.  For full details, see Section 2.2 of \citet{mollenhoff2019efficient}. In the Appendix, we provide simple R code that can be used to conduct this test.

{Second, we generate 25,000 datasets for each individual combination of 4 global parameters:

\begin{enumerate}
\item the population effect size ($\delta=\{0,0.01,0.02,\ldots,0.5\}$),
\item the sample size in each group ($n=\{50,100,250,500\}$),
\item the equivalence margin ($m=\{0.1,0.2,0.3\}$), and
\item the prior scale ($r=\{0.5/\sqrt{2}, 1/\sqrt{2}, 2/\sqrt{2}\}$).
\end{enumerate}  

\noindent To be clear, for each of the 1,836 ($=51 \times 4 \times 3 \times 3$) unique scenarios, we simulate 25,000 individual independent datasets.  

Unlike \citet{linde2020decisions}, we define $m$ as the \textit{unstandardized} equivalence margin.  We decided to define $m$ as the unstandardized margin rather than as the standardized margin so as to make things as simple as possible (the added simplicity will help us with discussing decision boundaries in Section 4) and because we are concerned about the improper use of equivalence tests defined with standardized margins; see \citet{campbell2020equivalence}.  (To be brief, \citet{lakens2017equivalence}’s suggestion that one may simply define the equivalence margin in terms of the observed standard deviation is technically incorrect.  Recall that a valid frequentist hypothesis cannot be defined in terms of the observed data. As such, if the equivalence margin is defined as a function of the observed standard deviation, then the equivalence test is invalid.)

For each dataset, we conduct four different procedures (the frequentist TOST, the frequentist optimal test, the Bayesian BF procedure, and the Bayesian HDI-ROPE procedure) and record:
\begin{itemize}
    \item the $p$-values obtained from the frequentist equivalence testing procedures,
    \item the BF obtained from the Bayes factor interval null procedure, and
    \item the maximum probability of the HDI at which the HDI-ROPE procedure will predict equivalence.
\end{itemize}

We specifically chose to conduct 25,000 simulation runs so as to keep computing time within a reasonable limit while also reducing the amount of Monte Carlo standard error to a negligible amount \footnote{ Note that since the prior-scale is irrelevant for the two frequentist procedures, we essentially obtain 75,000 simulations for the frequentist results for each of 612 ($=51 \times 4 \times 3$) unique scenarios.  We will consider only the first 25,000 of these (and disregard the remaining 50,000) when reporting the results so that the precision of the results for all methods is comparable.}}. Note that for computing a false positive rate that is truly $\alpha=0.50$, Monte Carlo SE will be approximately $0.003 \approx \sqrt{0.5(1-0.5)/25,000}$; see \cite{morris2019using}.  We ran all simulations using R based on the code provided by \cite{linde2020decisions} using parallel nodes of the Compute Canada cluster \citep{baldwin2012compute}.

Finally, we proceed by calibrating the Bayesian procedures (the HDI-ROPE and BF procedures) so that they maintain a predetermined maximum type 1 error rate of $\alpha$.  This is done by adjusting each procedure such that, for a given sample size, the proportion of equivalence predictions obtained is exactly $\alpha$ when the margin is equal to the population effect size, i.e., when  $m = \delta$.  The frequentist testing procedures will not be calibrated since, in theory, they should require no calibration as they are specifically designed to observe this property (at least asymptotically).  The scenario in which $m=\delta$ represents the boundary of the null hypothesis, $H_{0}: |\delta| \ge m$.   The calibration will therefore ensure that  whenever the null hypothesis is true (i.e., whenever $|\delta| \ge m$), the proportion of equivalence predictions will be no greater than $\alpha$.

 \cite{linde2020decisions}  seem to suggest that, when $\delta = m$ (at the boundary of the null hypothesis), it is desirable to have ``a proportion of equivalence decisions close to 0.5'' so that the test is ``an unbiased classifier [which] maximizes accuracy.''  We therefore consider, for the results of our simulation, the maximum type 1 error set with $\alpha=0.50$.  In addition, we will also report the results with $\alpha=0.05$, a common choice in frequentist analyses.  The ideal value for $\alpha$ will no doubt depend on the specific context of one’s analysis \citep{lakens2018justify}.  While it is true that $\alpha=0.5$ is uncommon in practice, we note that $\alpha=0.05$ is an equally arbitrary threshold which has come to prominence due primarily to a historical happenstance \citep{kennedy2019before}.

The BF procedure can be calibrated to be more or less conservative by setting a higher or lower BF decision threshold and/or by using a smaller/larger Cauchy prior scale.  However,  \citet{linde2020decisions} note that calibration by selecting a smaller/larger Cauchy prior scale ``is not advised.''  We proceed in the simulation study by calibrating the BF procedure by adjusting the BF decision threshold.  Calibration of the HDI-ROPE procedure can be done by selecting a smaller/larger prior scale and/or by adjusting the probability of the HDI.  We proceed in the simulation study by calibrating the HDI-ROPE procedure by adjusting the probability of the HDI.

\section{Results}
\begin{figure}
    \centering
    \includegraphics[width=14cm]{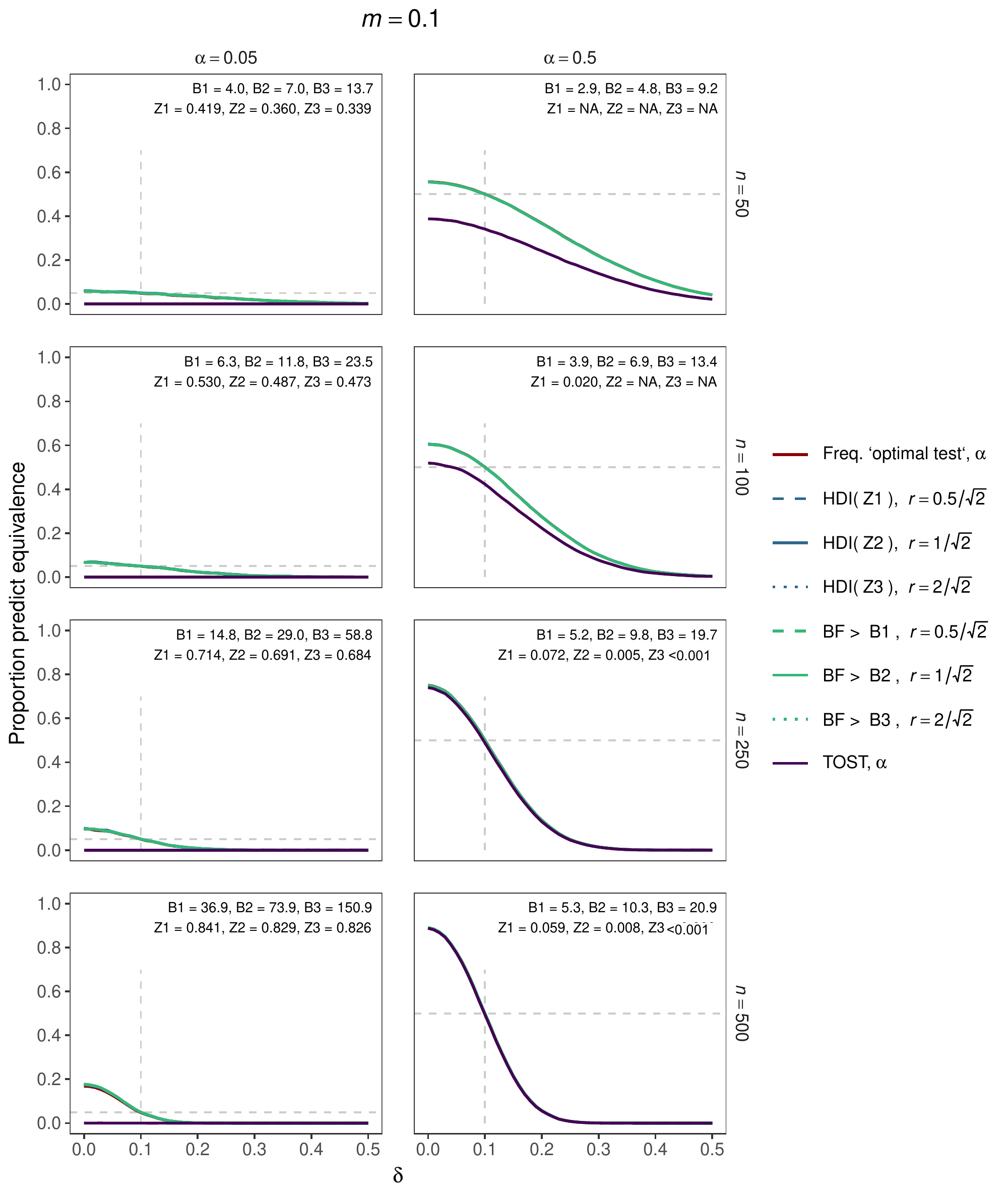}
 \caption{The proportion of equivalence predictions with an equivalence margin of $m = 0.1$ (vertical dashed line). Panels in the left-hand column correspond to results for $\alpha=0.05$ (horizontal dashed line), and panels in the right-hand column correspond to results for $\alpha=0.50$ (horizontal dashed line).  Each row of panels contains results for a different sample size ($n$). Colours denote the four different inferential approaches. Line types denote the three  different priors (for Bayesian procedures).  Each coloured line corresponds to simulation results from 25,000 simulation runs.  Predictions of equivalence are correct if the population effect size ($\delta$) lies within the equivalence interval (i.e., if $|\delta| < m$), whereas predictions of equivalence are incorrect if $\delta$ lies outside the equivalence interval (i.e., if $|\delta| \ge m$). Bayesian metrics are calibrated such that the proportion of equivalence predictions is exactly equal to $\alpha$ (horizontal dashed line) when $\delta = m$ (at the intersection of the horizontal and vertical dashed lines).  The calibration for the Bayesian procedures is specified by the Z1, Z2, and Z3 probability values for the HDI-ROPE procedure and by the B1, B2, and B3 decision threshold values for the BF procedure. Note that the frequentist `optimal test,' the BF procedures, and the HDI-ROPE procedures, all produce a very similar (almost identical) proportion of equivalence predictions and therefore the seven different curved lines (the red, blue and green lines) are not independently visible in any of the panels.  Also note that calibration of the HDI-ROPE procedure for five scenarios above (those with "NA" indicated for the Z1, Z2, and Z3 probability values) was not possible due to numerical limitations.    }
    \label{fig:my_label1}
\end{figure}

\begin{figure}
    \centering
   \includegraphics[width=14cm]{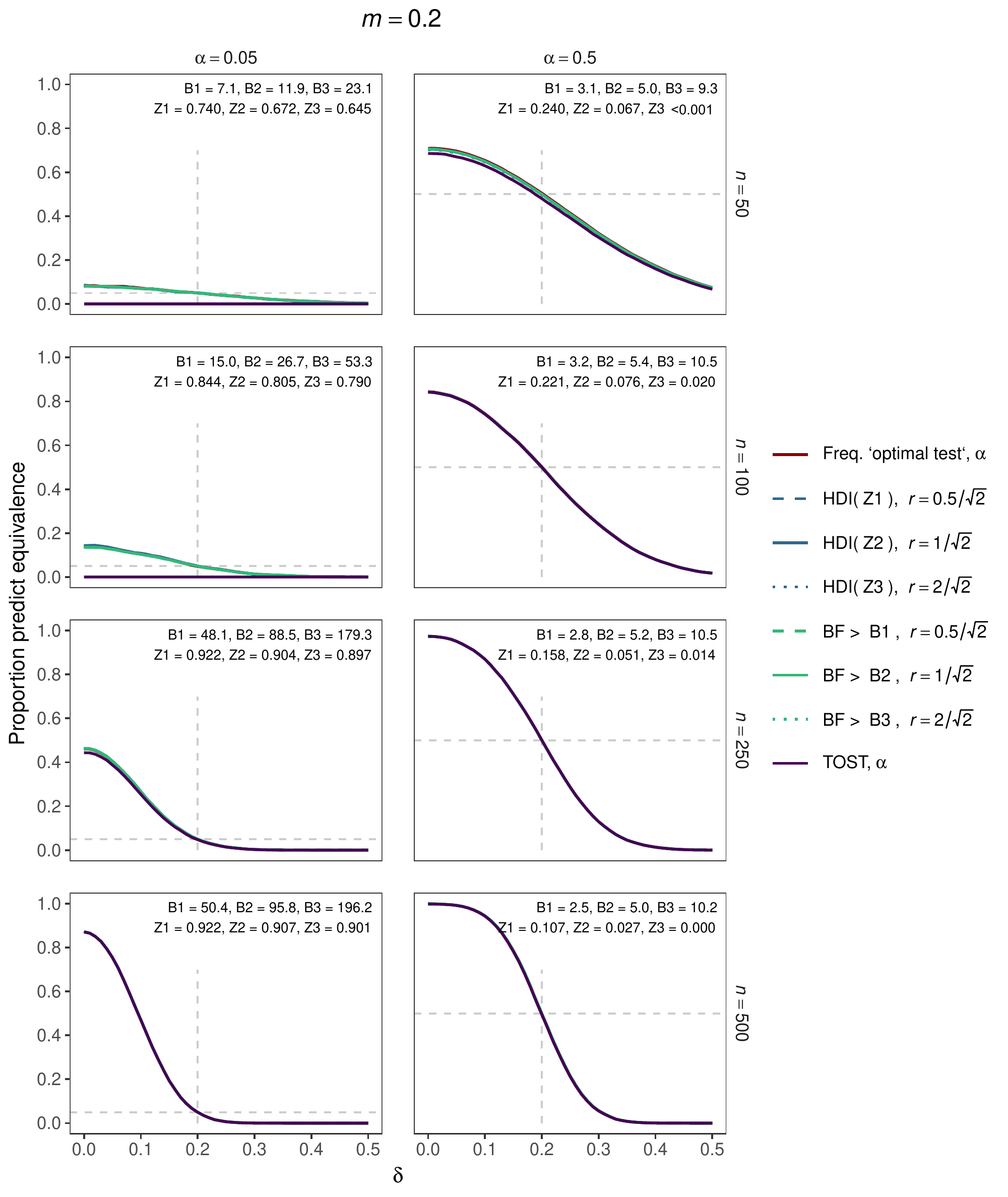}
 \caption{The proportion of equivalence predictions with an equivalence margin of $m = 0.2$ (vertical dashed line). Panels in the left-hand column correspond to results for $\alpha=0.05$ (horizontal dashed line), and panels in the right-hand column correspond to results for $\alpha=0.50$ (horizontal dashed line).  Each row of panels contains results for a different sample size ($n$). Colours denote the four different inferential approaches. Line types denote the three different priors (for Bayesian procedures).  Each coloured line corresponds to simulation results from 25,000 simulation runs.  Predictions of equivalence are correct if the population effect size ($\delta$) lies within the equivalence interval (i.e., if $|\delta| < m$), whereas predictions of equivalence are incorrect if $\delta$ lies outside the equivalence interval (i.e., if $|\delta| \ge m$). Bayesian metrics are calibrated such that the proportion of equivalence predictions is exactly $\alpha=0.5$ (horizontal dashed line) when $\delta = m$ (at the intersection of the horizontal and vertical dashed lines).  The calibration for the Bayesian procedures is specified by the Z1, Z2, and Z3 probability values for the HDI-ROPE procedure and by the B1, B2, and B3 decision threshold values for the BF procedure. Note that all of the procedures (with the exception of the TOST procedure for $\alpha=0.05$ and $n=50, 100$) produce a very similar (almost identical) proportion of equivalence predictions and therefore the eight different curved lines (the purple, blue, green, and maroon lines) are not independently visible.  }
   \label{fig:my_label2}
\end{figure}

\begin{figure}
    \centering
    \includegraphics[width=14cm]{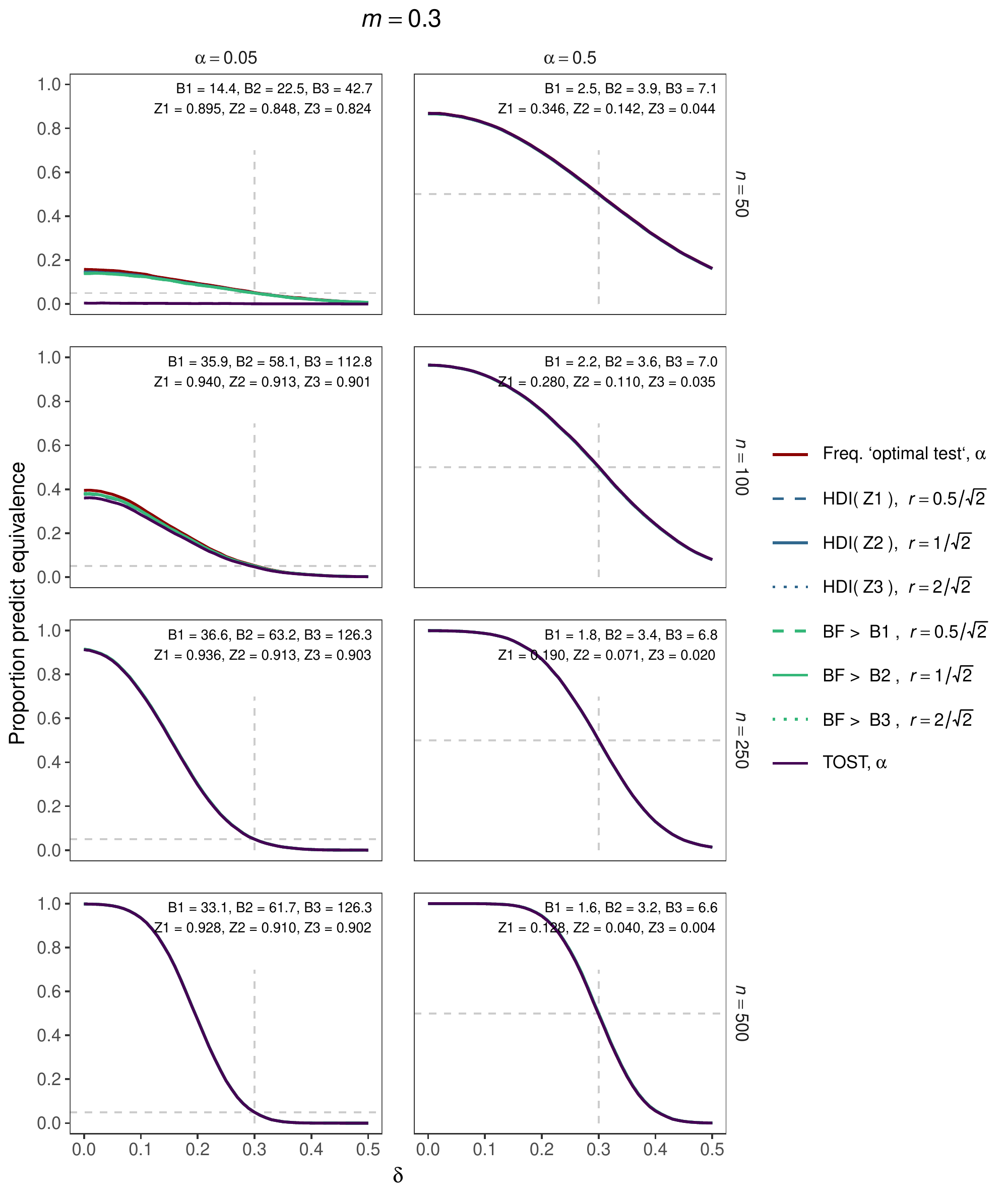}
 \caption{The proportion of equivalence predictions with an equivalence margin of $m = 0.3$ (vertical dashed line). Panels in the left-hand column correspond to results for $\alpha=0.05$ (horizontal dashed line), and panels in the right-hand column correspond to results for $\alpha=0.50$ (horizontal dashed line).  Each row of panels contains results for a different sample size ($n$).  Colours denote the four different inferential approaches. Line types denote the three  different priors (for Bayesian procedures).  Each coloured line corresponds to simulation results from 25,000 simulation runs.  Predictions of equivalence are correct if the population effect size ($\delta$) lies within the equivalence interval (i.e., if $|\delta| < m$), whereas predictions of equivalence are incorrect if $\delta$ lies outside the equivalence interval (i.e., if $|\delta| \ge m$). Bayesian metrics are calibrated such that the proportion of equivalence predictions is exactly $\alpha=0.5$ (horizontal dashed line) when $\delta = m$ (at the intersection of the horizontal and vertical dashed lines).  The calibration for the Bayesian procedures is specified by the Z1, Z2, and Z3 probability values for the HDI-ROPE procedure and by the B1, B2, and B3 decision threshold values for the BF procedure. Note that all of the procedures (with the exception of the TOST procedure for $\alpha=0.05$ and $n=50$) produce a very similar (almost identical) proportion of equivalence predictions and therefore the eight different curved lines (the purple, blue, green, and maroon lines) are not independently visible. }
    \label{fig:my_label3}
\end{figure}

The results of the simulation study with $\alpha=0.05$ are shown in the left-hand panels of Figures \ref{fig:my_label1}, \ref{fig:my_label2}, and \ref{fig:my_label3}, for equivalence margins of $m = 0.1$, $m = 0.2$, and $m = 0.3$ respectively. The results of the simulation study with $\alpha=0.5$ are shown in the right-hand panels.  We have three main comments.

First and foremost, note that, when calibrated, the frequentist optimal test, the frequentist-calibrated HDI-ROPE and the frequentist-calibrated BF all display an almost identical probability of predicting equivalence for all values of $n$, $m$, $\delta$, $r$, and $\alpha$.  This suggests that, for the vast majority of scenarios, regardless of which of these approaches is used for analysis, the result could be made the same by adopting the same calibration.  With regards to the TOST, our results are similar to those reported by \citet{linde2020decisions}: the TOST, when $\alpha=0.05$, has little power to establish equivalence when $n$ and/or $m$ are small.  With $\alpha=0.5$, the TOST appears to be overly conservative when $n$ and $m$ are small (see top panels in Figure \ref{fig:my_label1}).  Note that we were unable to adequately calibrate the HDI-ROPE procedure for five small sample size scenarios with $\alpha=0.5$ and $m=0.1$ due to insufficient numerical accuracy of our results.  In these five scenarios (which can be identified as those with "NA" indicated for the Z1, Z2, and Z3 values in Figure \ref{fig:my_label1}), the probability of the HDI that was required for calibration was less than 0.0001 and impossible to determine with sufficient accuracy.

Second, when calibrated to obtain a specific predetermined maximum type 1 error rate, the Bayesian procedures appear to operate identically regardless of one's choice of prior-scale.  This is immediately obvious in Figures \ref{fig:my_label1}-\ref{fig:my_label3}: the three blue (green) lines -dashed ($r=0.5/\sqrt{2}$), solid ($r=1/\sqrt{2}$), and dotted ($r=2/\sqrt{2}$)- for the HDI-ROPE (the BF), are indistinguishable from one another.  This suggests that choosing a smaller or larger prior-scale is essentially irrelevant, at least from a frequentist perspective, for a calibrated Bayesian procedure.

Finally, with $\alpha=0.05$, the probability values required to calibrate the HDI-ROPE procedure range from 0.34 to 0.94 and appear to increase with decreasing values of $r$, with increasing values of $m$, and with increasing values of $n$.  With $\alpha=0.5$, the probability values required to calibrate the HDI-ROPE procedure are much much lower (and in some cases impossible to determine with sufficient accuracy).  We suspect that as $n$ increases, the probability value required to calibrate the HDI-ROPE procedure will trend towards $(1-2\alpha)$ since the credible interval will closely approximate the confidence interval with a sufficiently large $n$.  The BF decision thresholds required to calibrate the BF procedure with $\alpha=0.05$ range from 4.0 to 196.2, and with $\alpha=0.5$, range from 1.6 to 20.9.  The BF decision thresholds appear to increase with increasing values of $r$, with increasing values of $n$, and with decreasing values of $\alpha$.

\section{Discussion}

The simulation study results suggest that the Bayes Factor, HDI-ROPE, and frequentist equivalence tests can all be reverse engineered to achieve the same operating characteristics as one another.  While the simulation study is limited to two-sample normally distributed data, we suspect that a similar conclusion could be made in other scenarios.  In order to better understand, it is useful to consider the sufficient statistics required for each procedure.

For a given value of the observed absolute difference in sample means ($|\bar{X}_{1} - \bar{X}_{2}|$), a given value of the observed pooled standard deviation  ($\hat{\sigma}_{P}$), and a fixed sample size ($n$) and margin ($m$), there is a single unique $p$-value that one will obtain from the frequentist optimal test,  a single unique $p$-value one will obtain from the TOST, a single unique BF one will obtain from the BF procedure (with a given prior-scale), and a single unique probability at which the HDI-ROPE procedure (with a given prior-scale) will predict equivalence.  As such, we can easily determine a 2-dimensional (dimension 1: $|\bar{X}_{1} - \bar{X}_{2}|$; dimension 2: $\hat{\sigma}_{P}$) decision threshold for each of the four procedures.  

\begin{figure}
    \centering
    \includegraphics[width=14cm]{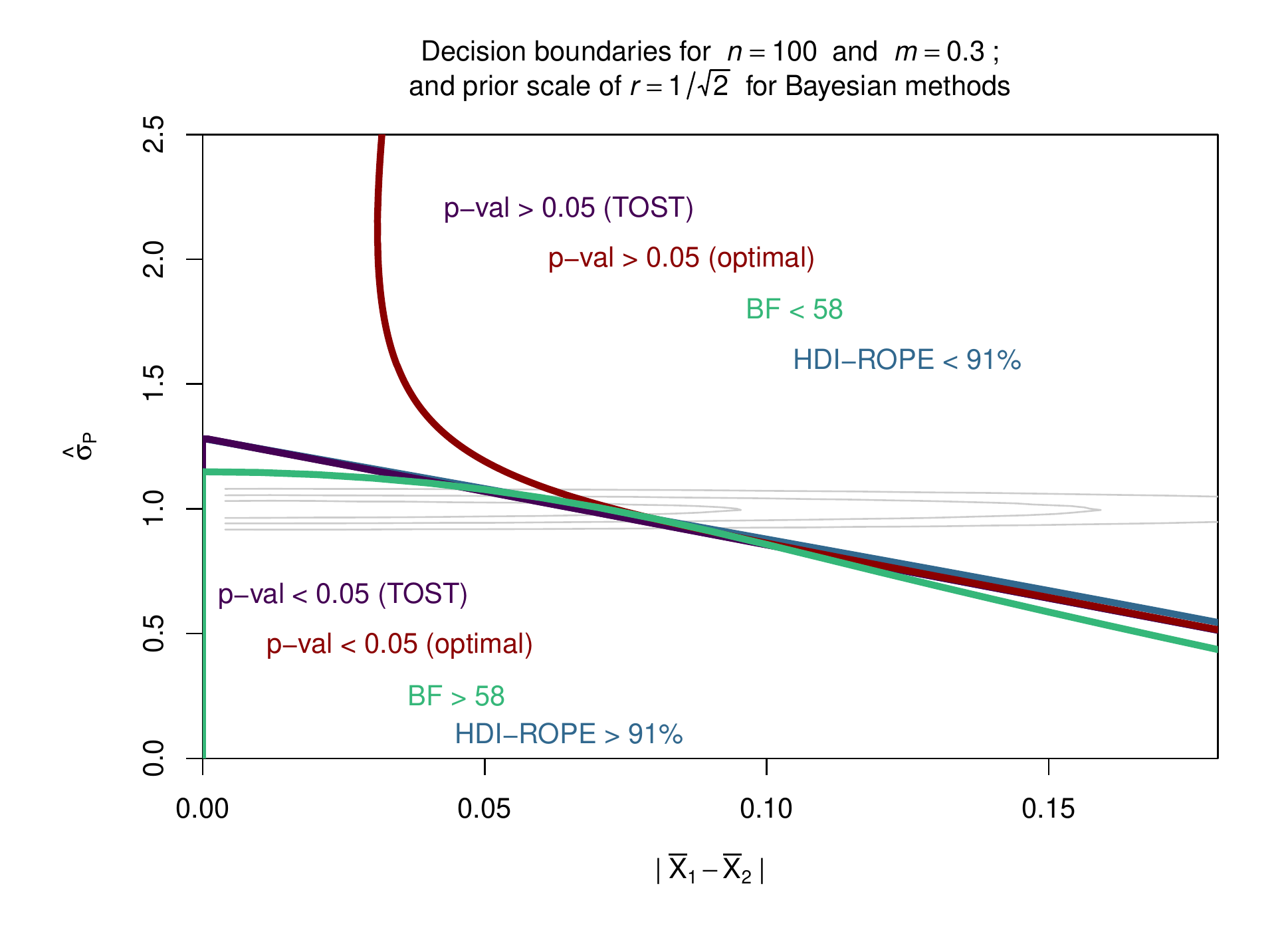}
    \caption{The different decision boundary lines, in terms of $|\bar{X}_{1}-\bar{X}_{2}|$ and $\hat{\sigma}_{P}$, for each of the four procedures for $n=100$ and $m=0.3$, and with a prior-scale of $r=1/\sqrt{2}$ for the Bayesian procedures.  The grey contour lines correspond to the two-dimensional density of the distribution of observed $|\bar{X}_{1} - \bar{X}_{2}|$ and $\hat{\sigma}_{P}$ values obtained from simulating ten million independent datasets from $X_{1} \sim N(0,1)$ and $X_{2} \sim N(0,1)$  (with $n=100$).}
    \label{fig:boundary}
\end{figure}

Figure \ref{fig:boundary} plots the different decision boundary lines for each of the four procedures for $\alpha=0.05$, $n=100$ and $m=0.3$, and with a prior-scale of $r=1/\sqrt{2}$ for the Bayesian procedures (these calibrated based on the simulation study results which gave $B1=58.1$, $Z1=0.913$).  For reference, we have overlaid on this plot the two-dimensional density of the distribution of observed $|\bar{X}_{1} - \bar{X}_{2}|$ and $\hat{\sigma}_{P}$ values obtained from simulating ten million independent datasets from $X_{1} \sim N(0,1)$ and $X_{2} \sim N(0,1)$.  From this figure, we conclude that, with regards to where the vast majority of the data will be observed (i.e., the area outlined by the grey contour lines), the decision boundaries for all four procedures are nearly identical. 

\begin{figure}
    \centering
    \includegraphics[width=14cm]{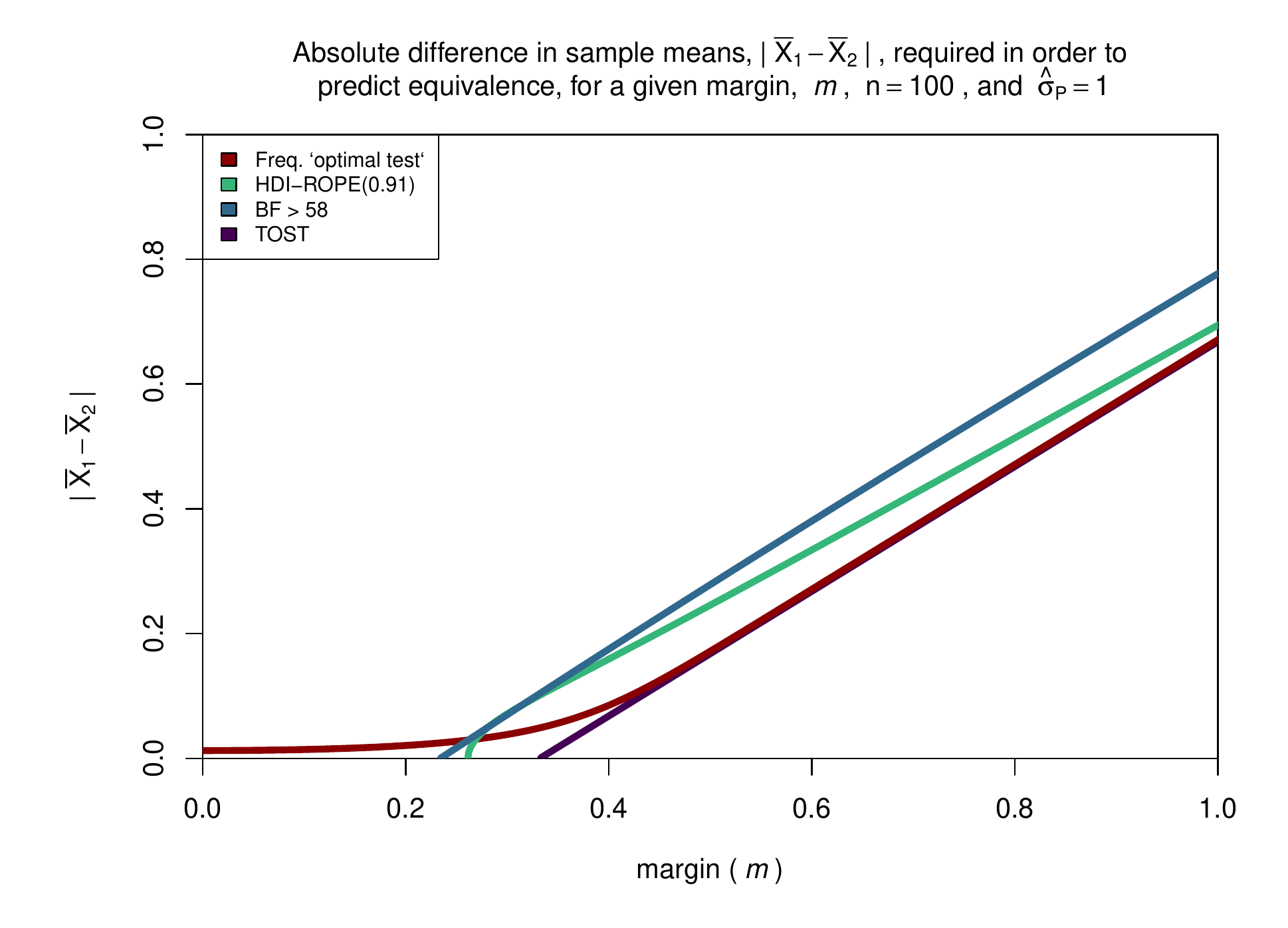}
    \caption{The value of $|\bar{X}_{1} - \bar{X}_{2}|$ that would be required in order to predict equivalence for a range of values of $m$, $n=100$, and for a given $\hat{\sigma}_{P}=1$.  Priors for the two Bayesian procedures are set with a prior-scale of $r=1/\sqrt{2}$.  The frequentist procedures are calibrated with $\alpha=0.05$.}
    \label{fig:margin to delta}
\end{figure}

Figure \ref{fig:margin to delta} plots the maximum value of $|\bar{X}_{1} - \bar{X}_{2}|$ that allows one to predict equivalence for a range of $m$, and for $n=100$ and a given $\hat{\sigma}_{P}=1$.  We set a  prior-scale of $r=1/\sqrt{2}$ for the Bayesian procedures which are calibrated based on the results of the simulation study.  The frequentist procedures are calibrated with $\alpha=0.05$.  For the HDI-ROPE, the BF and the TOST, note that there are values of $m$ for which one cannot predict equivalence, regardless of the value of $|\bar{X}_{1} - \bar{X}_{2}|$.  However, for the optimal frequentist test, there will always be a value for $|\bar{X}_{1} - \bar{X}_{2}|$ small enough to predict equivalence no matter how small the value of $m$.

\begin{figure}
    \centering
    \includegraphics[width=14cm]{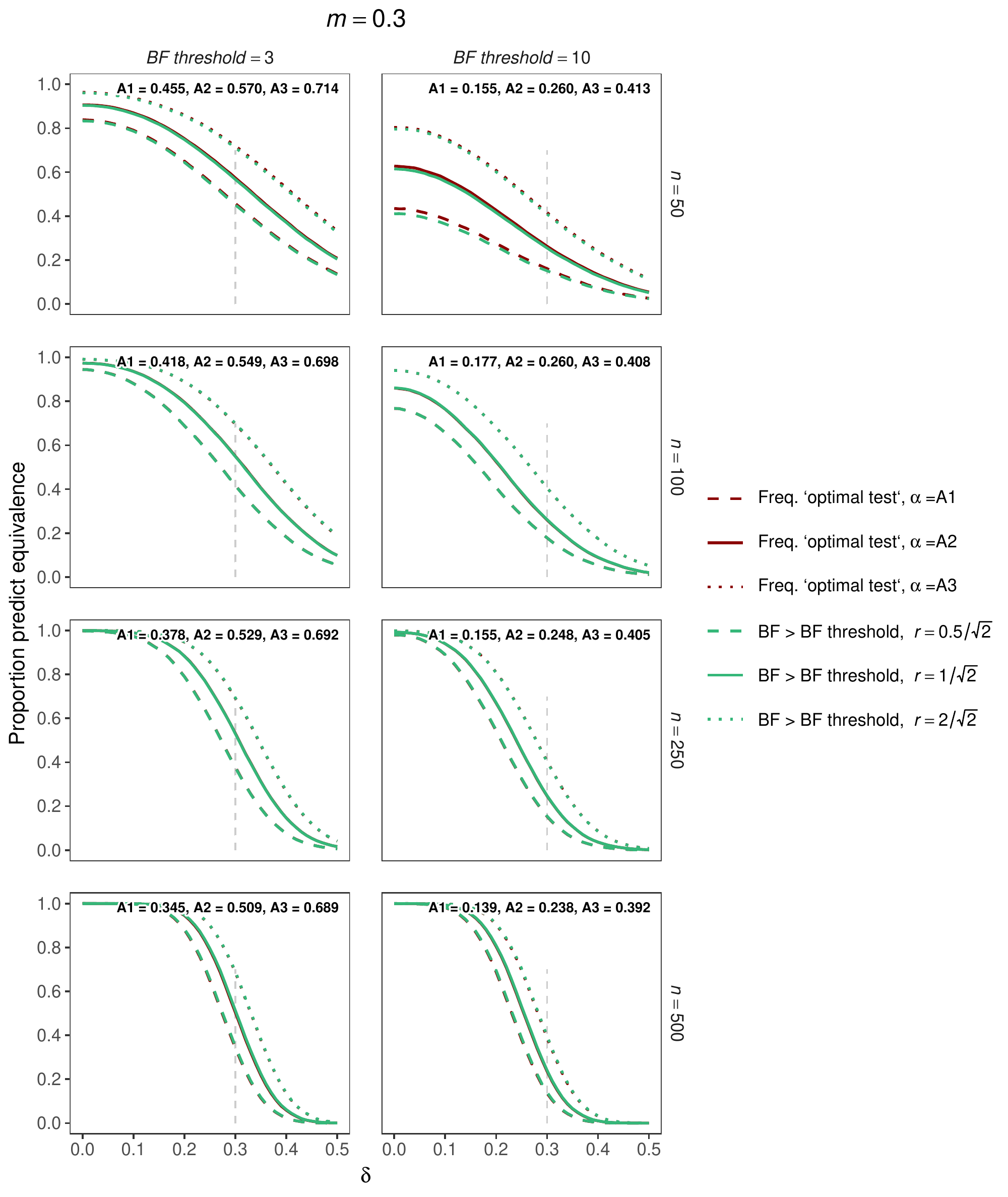}
    \caption{  The proportion of equivalence predictions with an equivalence margin of $m = 0.3$ (vertical dashed line) and with the $\alpha$ level chosen for each scenario so that, at $m=\delta$, the proportion of equivalence predictions made with the optimal frequentist test matches the proportion made with the Bayes factor test with a BF threshold of 3 (panels in left-hand column) and a BF threshold of 10 (panels in right-hand column). Panels in each row contain results for a different sample size ($n$). Colours denote the two different inferential approaches. Line types denote the three different priors (for the Bayesian procedure).  Each coloured line corresponds to simulation results from 25,000 simulation runs.  Predictions of equivalence are correct if the population effect size ($\delta$) lies within the equivalence interval (i.e., if $|\delta| < m$), whereas predictions of equivalence are incorrect if $\delta$ lies outside the equivalence interval (i.e., if $|\delta| \ge m$).  The calibration for the frequentist procedure is specified by the A1, A2, and A3 values for $\alpha$. Note that the frequentist `optimal test,' and the BF procedures produce a very similar (almost identical) proportion of equivalence predictions and therefore the red and green lines are not independently visible.}
    \label{fig:freq3m03}
\end{figure}

 In summary, it appears that, for fixed $n$, we can almost exactly reproduce a frequentist test at a given level $\alpha$ by adopting a Bayesian procedure and reverse-engineering it (by determining necessary values for either the $BF$ threshold or the probability of the HDI).   This street, however, is open to two-way traffic. That is, we should also be able to mimic a particular Bayesian test with a frequentist test, by reverse-engineering the $\alpha$ level.  See Figure \ref{fig:freq3m03} in which our simulation study results for $m=3$ are plotted again but with the $\alpha$ level chosen for each scenario so that, at $\delta=m$, the proportion of equivalence predictions made with the optimal frequentist test matches the proportion made with the Bayes factor test, with thresholds of 3 and 10.  The notion of ``two-way traffic'' becomes clear when we are reminded that the Bayes factor has an underlying principle of its own, albeit one that is much less publicized than the frequentist control of the type 1 error rate.
 

 If we imagine repeated sampling of datasets, with the underlying parameters themselves being {\em different} from draw to draw, then the average performance of Bayes factor testing has a decision-theoretic optimality. Specifically, if the parameters are drawn from the overall prior distribution, which itself is a mixture of the “under null” and “under alternative” prior distributions, then the BF procedure minimizes the average loss, often referred to as the Bayes' risk \citep{berger2013statistical}.
 
 Consider the simplest case of 50\% prior weight on each of the null and alternative, and a loss function that weights type 1 and type 2 errors equally. Then the procedure which selects the null or alternative by comparing the Bayes' factor to an evidence threshold of 1 minimizes the probability of a selection error (with respect to the particular sense of repeated sampling described above). More generally, with prior probability $1-q$ on the null and $q$ on the alternative, and a type 1 error deemed to be $k$ times as damaging as a type 2 error, the average loss is minimized by basing selection on comparison of the BF to a threshold of $k(1-q)/q$. 
 
 So Bayes factor testing indeed has an underlying premise and interpretation – it just happens to differ from the frequentist principle of minimizing the probability of a type 2 error subject to an upper-bound on the maximum probability of a type 1 error; see \citet{berger2013statistical}.  Coming back to the two-way traffic then, if one desires to carry out Bayesian testing as is rooted in the interpretation above, then for fixed $n$, one could reverse-engineer a value of $\alpha$ such that the frequentist test would almost exactly do the job, in terms of reproducing the decision boundary.  Indeed this is what we see in Figure \ref{fig:freq3m03}.  
 
 Finally, to be clear, a Bayesian, in principle, should not be concerned with minimizing the type 2 error, given a fixed upper-bound on the type 1 error rate.  And conversely, a frequentist, in principle, should not be concerned with minimizing the Bayes’ risk.  However, in practice, there is nothing preventing a Bayesian from using a frequentist test calibrated in such a way so as to minimize the Bayes’ risk, and nothing preventing a frequentist from using a Bayes factor calibrated in such a way so as to control the type 1 error.  There is, however, a fundamental difference between the Bayesian and the frequentist when it comes to how to consider the sample size.  The frequentist will maintain the same value for $\alpha$ regardless of $n$, whereas the Bayesian will adjust $\alpha$ depending on $n$; see \citet{wagenmakers2021history}.  Indeed, it is only after observing how a researcher treats two samples of different sample sizes, that one could reliably determine whether the researcher is acting as a frequentist or as a Bayesian.

\section{Conclusion}

In general any advocating for frequentist testing as better or worse than Bayesian testing in terms of empirical findings seems dubious at best. If you decide on which underlying principle you want to subscribe to in tackling a given problem, then the method follows naturally. And particularly bearing in mind that either procedure can be reverse-engineered from the other (at least approximately), as we have shown, trying to use empirical performance to argue for one over the other seems like tilting at windmills.  This being said, it is crucial to understand how a given statistical test, be it either frequentist or Bayesian, operates under different circumstances.  Understanding a statistical procedure's operating characteristics is key to ensuring its proper use, and, perhaps more importantly, key to avoiding its misuse.

Recall, as an example of a misused statistical procedure, the controversial method of ``magnitude-based inference'' (MBI) \citep{barker2008inference}.  While rarely used or even  acknowledged in other fields, MBI became widely popular in sports medicine research.   The supposedly  ``philosophically and statistically distinct'' \citep{batterham2015} statistical procedure was poorly understood and led countless sports medicine researchers to unreliable and altogether erroneous conclusions.  Only once the operating characteristics of MBI were better understood \citep{sainani2019magnitude, sainani2018problem}  were researchers advised to avoid using it for their analyses \citep{lohse2020systematic}.  Unfortunately, by that point, much damage had already been done to the field of sports medicine research.

 All too often, non-significance (e.g., $p > 0.05$), or a combination of both non-significance and supposed high statistical power, is used as the basis to claim the lack of a meaningful effect.  This approach is logically flawed.  As the saying goes, ``absence of evidence is not evidence of absence'' ~\citep{hartung1983absence, altman1995statistics}.  Researchers should instead use one of the tools available for equivalence testing.  Based on our simulation study, we determined that frequentist equivalence tests, the Bayes factor, and the HDI-ROPE can all be calibrated to be roughly equivalent in terms of their power to detect the lack of a meaningful effect.

\citet{linde2020decisions} concluded ``that the BF approach is particularly useful for research with small sample sizes.''  Our simulation results suggest otherwise.  We observed nothing ``particularly useful about the BF approach.''   With this in mind, we recommend that researchers, if they can properly calibrate \textit{and communicate} their results, use whatever approach suits them best.  A potential advantage with frequentist tests is that they are widely used and well understood in fields outside of psychology \citep{wellek2010testing, jones1996trials}.  The same cannot be said for the HDI-ROPE or the Bayes factor interval null procedures.

If the Bayes factor interval null procedure is used for predicting equivalence with standard BF decision thresholds such as $BF_{thr}=3$ or $BF_{thr}=10$ (i.e., used without frequentist calibration), one should expect to see a very high false positive rate.  Indeed, \cite{linde2020decisions} observed false positive rates higher than 60\% for both $BF_{thr}=3$ and $BF_{thr}=10$ when $m=0.1$ and $r=\sqrt{2}/2$.  In contrast, if the HDI-ROPE procedure is used with a standard 95\% HDI (i.e., used without frequentist calibration), one should expect to see a very low false positive rate, well bellow 5\%.

With small sample sizes, the TOST procedure may indeed ``have no discriminatory power and result in a foregone decision for non-equivalence'' \citep{linde2020decisions}.  For this reason, researchers are advised to use the so-called ``optimal test'' based on the folded Normal distribution \citep{romano2005optimal} rather than the TOST procedure when sample sizes are very small.  Note that, regardless of which frequentist testing procedure is used, researchers must be careful to select an appropriate equivalence margin.  This is often easier said than done; see \citet{campbell2018make}.

Finally, given a number of different procedures that, when calibrated, are essentially identical in terms of their statistical power, one might question why some researchers will prefer one approach over another \citep{andrews2013prior, dienes2014using}.  To answer this, we must recognize that statistics are not entirely defined by statistical power metrics and their operating characteristics.  Indeed, it is important to understand that statistics are, as \cite{kasy2019selective} wisely note, ``a \textit{social} process of communication and collective learning that involves many different actors with differences in knowledge and expertise, different objectives, and constraints on their attention and time, who engage in strategic behavior."

\bibliography{references}

\pagebreak

\section{Appendix}
Below is R code for the so-called ``optimal'' frequentist equivalence test based on the folded-Normal distribution; see details in Section 2.2 of \citet{mollenhoff2019efficient}.

\begin{footnotesize}
\begin{verbatim}
## optim_equiv: a function for two-sample equivalence
## testing.  Produces both TOST p-val and optimal test p-val
optim_equiv <- function(sample1, sample2, margin) {
require("VGAM")
n1 <- length(sample1); n2 <- length(sample2)
s2_1 <- sd(sample1)^2; s2_2 <- sd(sample2)^2
s_P = sqrt(( ((n1 - 1) * s2_1) + 
	         	((n2 - 1) * s2_2) )/(n1 + n2 - 2))
xbar1 <- mean(sample1); xbar2 <- mean(sample2)
se.diff <- (s_P*sqrt(1/n1 + 1/n2))
t_1 <- (xbar1 - xbar2 - (-margin))/se.diff
t_2 <- (xbar1 - xbar2 - (margin))/se.diff
pval1 <- 1 - pt(t_1, n1 + n2 - 2)
pval2 <- 1 - pt(t_2, n1 + n2 - 2, lower = FALSE)
tost_pval <- max(c(pval1, pval2))
optimal_equiv <- function(x){	abs(xbar1 - xbar2) - qfoldnorm(x, margin, se.diff)	}
optim_pval <- NA
if(is.na(optim_pval)){
	tryCatch({optim_pval <- uniroot(optimal_equiv, 
			c(0, (1 - 1/10e15)), tol = 0.0001)$root
	}, error=function(e){})}

return(c(tost = tost_pval, optim = optim_pval))}

# Examples:
set.seed(123)
optim_equiv(rnorm(100), rnorm(260), margin = 0.4)
#        tost       optim 
# 0.003542515 0.003349803 
optim_equiv(rnorm(40), rnorm(26), margin = 0.4)
#       tost      optim 
# 0.05371685 0.01259863     
\end{verbatim}
\end{footnotesize}
\end{document}